\documentclass[oneside]{article}

\usepackage{epsfig}
\usepackage{amssymb,color}
\usepackage{framed}
\usepackage{epsf}
\usepackage{algorithm, algorithmic}
\usepackage{booktabs}
\usepackage{multirow}
\usepackage{graphics,epsfig}
\usepackage[normalem]{ulem}
\usepackage{url}

\newcommand{\qed}{\mbox{}\hspace*{\fill}\nolinebreak\mbox{$\rule{0.6em}{0.6em}$}
}

\definecolor{gray}{rgb}{0.5,0.5,0.5}

\newtheorem{theorem}{Theorem}

\newtheorem{question}[theorem]{Question}

\begin{document}
\title{Can Knowledge be preserved in the long run?}

\iftrue
\author{
Rina Panigrahy \\
Microsoft Research, Mountain View, CA \\
rina@microsoft.com
}
\fi

\maketitle

\newcommand{\Omg}{$\Omega$}

\begin{abstract}
Can (scientific) knowledge be reliably preserved over the long term? We have today very efficient and reliable methods to encode, store and retrieve data in  a storage medium that is fault tolerant against many types of failures. But does this guarantee -- or does it even seem likely -- that all knowledge can be preserved over thousands of years and beyond? History shows that many types of knowledge that were known before have been lost.
We observe that  the nature of stored and communicated information and the way it is interpreted is such that it always tends to decay and therefore must lost eventually in the long term. The likely fundamental conclusion is that knowledge cannot be reliably preserved indefinitely.
\end{abstract}

\section{The Problem of  Preserving Knowledge}

The question that we will study is  (it may seem a bit philosophical but really it is a pragmatic one)

\begin{question}
Is there a mechanism to preserve knowledge in the long term?

OR   is it the case that

\noindent Over the long term a large fraction of knowledge must be `provably' lost.
\end{question}

Our main thesis is that preserving knowledge may not be easy and the latter part of the question seems to be more likely to be true. This would mean that (any)knowledge must necessarily be lost in the long term and this would be a fundamental fact about nature -- knowledge, like material objects,  has only a limited existence. At the very least, we will argue that even if knowledge can be preserved in the long term, achieving this is by no means easy.
By knowledge we mean let's say all the scientific papers ever written so far or all the philosophical texts written so far.  So our question is whether such knowledge can be preserved over a long time -- 100's 1000's 10000's  years.
Knowledge is information that enables us to do things better, to improve our lives, or to improve our understanding of ourselves and the world we live in, or of abstract concepts inspired from models of processes in the world.
To have knowledge means to have the ability to make use of the knowledge when required - for example having knowledge of a certain technology is measured by  the ability to build the technology -- having a book which 
describes the technology, but by reading which one is not able to build the technology is not considered complete knowledge of the technology. Similarly to lose a piece of knowledge is to lose the ability to take advantage of that knowledge.

This question is not new and has been (or essentially the same question in spirit)  asked many times before~\cite{Mach,Mach2, Madhu,Orig,Summa}. The field of coding theory and error correcting codes~\cite{madhucoding, infotheory} deals with methods of representing, storing and transmitting data and the underlying information so as to protect it from errors and corruption -- this question has the same goal -- to be able to preserve information over the long term.  Coding theory shows that in principle if we encode data using error correcting codes, self-correct it from time to time, then we should not lose the data for an exponential amount of time, assuming  a constant rate of error.  On the other hand, from a social science and economics perspective, Machlup~\cite{Mach}, in 1962, presented a detailed analysis of the knowledge industry in the US -- he coined the phrase `half life of knowledge' comparing knowledge to a radioactive material that decays over time.  The challenge of preserving existing books and papers effectively is voiced in several blogs and papers that talk about the growing complexity of human knowledge and the  difficulty of managing it~\cite{queens,timebomb,preservationdigital,knowledgemanage}. Sudan and Juba~\cite{Madhu} attempt to find ways to communicate mathematical knowledge using a {\em universal} language -- one that may be used for transmitting knowledge across civilizations even if they don't speak the same language. In a science fiction work,  Summa Technologiae~\cite{Summa}, Stanisław Lemm presents hypothetical scenarios on the future of technology and scientific knowledge -- the book was written in 1962 and many of the technologies he envisions seem to be real technologies of today -- some parts of the book deal with growing complexity of knowledge and its consequences.

In the next section we will argue that there are many challenges in preserving knowledge. Next in section~\ref{howloss} we will describe several ways in which knowledge gets lost. Next (section~\ref{forces}) we will attempt to understand the nature and the strength of the forces that destroy and produce knowledge. Finally, before concluding, we will make an observation (section~\ref{illusion}) that it is easy to lose knowledge without realizing it, and the implications of this observation to the notion of growing knowledge.

\section{The Challenges in  Preserving Knowledge}

Data can be preserved for an 
exponentially long time  using error correcting codes by constantly auto-correcting it.  But is this really true?
Just preserving the bits is not enough; there is always some program or  machine that extracts and {\em interprets} the bits of data.
What if the program is lost? So  maybe you can replicate the program or store it reliably. But what
about the compiler of the program? What about the hardware where the compiler runs?
Don't these hardware keep getting upgraded? Hardware becomes obsolete;
software infrastructures become obsolete.

What's more the instruction manuals for operating these machines are
written in a human readable language. These manuals may only be
understood by a few experts. If the experts are not retrained over
time, the information in the manual may be lost.

The data may represent textual information that may be written in some language.
Over 10000's years languages may change, books may perish, disks may
perish. If meanings of words change, rules of grammar change, what manuals would be readable, what texts would be
comprehensible, what wisdoms would be accessible? (It is not easy to read books that were
written several hundred years ago; yes, even the ones written in English) If data represents
images or videos, then the encoding formats  may become obsolete
\footnote{Is preserving images easier than preserving text as updating the image format may be automated, but
updating the words and sentences in a text with changing grammar rules may be harder?}.

The amount of `knowledge' being generated is increasing  (at least) linearly with
time. How can we hope to `maintain' an increasing body of
knowledge.
May be we can realistically only retain a small fraction of this increasing body.
It's like having a limited amount of cache of memory with new
`knowledge' constantly arriving overwriting on some older piece of knowledge that is
lost forever. If all  history is knowledge of some kind, can we ever
keep all of history? But history can
have an `infinite' resolution. We can't keep track of every second or
millisecond of history at the level of every person. Already we need to
sacrifice the resolution
of history. Perhaps we can only track history to a certain limited
length of time and to a certain limited depth. If this is the case
with history, this has to be the case
with the `history of knowledge'. History shows that much (perhaps most)
knowledge including what was considered the most precious (say philosophy, religions)
were lost and needed to be rediscovered.

Even if we have an exponential amount of disk space to store data, there are still
problems with retaining knowledge. How would you index it? How would
you access it? Would a Wikipedia like link structure help? Who would maintain the
ever growing number of pages and links? 
For knowledge to be lost, it is not necessary that the books/papers describing the knowledge be destroyed. Just as a book that is placed in the wrong catalogue section in the library is very hard to find again and is {\em essentially} lost,  similarly if some knowledge is described in  a web page but the link to that page is hard to find because say it is embedded deep inside some other webpage or say it doesn't quite surface in the initial many search results, then that web page is essentially lost -- or at least hard to find when needed. Just a flood of knowledge of different types may cause some piece of knowledge to become obscure and lost in the maze.
There may also be a trade-off between the  maintainability of  a storage medium  and its capacity. For example, text stored on stone tablets may be more durable than that stored it on disks; but it suffers from
low capacity. Also decoding the words from the bits stored on a disk requires more complex operations than reading from a stone tablet. While technology gives us storage mediums of increasing capacity, it also comes with increased complexity
of maintaining and decoding the information and thus in a sense lower durability.

Thus could it be the case that our knowledge is always growing and we can only afford
to retain a `constant' amount that we can handle? If so, this would mean we should
be very careful in choosing
which knowledge to retain and preserve over time as there is a limited
amount of effort available in making sure it is not lost possibly for
ever. But then `with high probability' won't we by chance lose the sense
of importance for some useful knowledge and disregard it and eventually lose
it?

Knowledge needs to be maintained, retranslated, re-understood,
rewritten even rediscovered. Can knowledge always be made succinct without losing it? May
be a well written body of
knowledge is one that is concise -- one that is able to add new
knowledge in a way that compresses and absorbs much of the old
knowledge in a new concise form.

\section{How can we let Knowledge slip into Oblivion?}\label{howloss}

So how exactly does knowledge disappear from our hands? If we lose it,  this must happen without we noticing it -- for if we notice it slipping away we wouldn't let it go -- or if we let it go we didn't think of it as important  knowledge; that is we didn't {\em recognize} it as knowledge. How can we ever let go of useful knowledge?  To understand this we need to understand what makes us treasure knowledge. Sometimes we treasure knowledge because it is directly useful; it makes our lives easier and better -- such as knowledge of science and technology. Sometimes it gives us great insights -- helps us understand the world we live in, understand ourselves, understand abstract concepts that we use to model the world -- such as astronomy, pure (may not be applied) physics, pure  mathematics, philosophy, morality.

Some types of knowledge -- say about a piece of technology -- can fade if it is not put to practice from time to time, if it is not `working' knowledge. Even if there is a book describing the technology, by simply not executing the knowledge, one can forget the exact details of how produce the technology. Any person who has written a computer program without documenting it well is familiar with the experience of having to struggle through ones memory in deciphering what the program is doing or how to exactly execute it with the right parameters if one hasn't used it in a long time. Sometimes knowledge is lost with a loss of a civilization, sometimes with a change of culture that values one type of knowledge over another, sometimes it is lost in war where a defeated society's knowledge is lost from prominence in the world stage. Sometimes a technology is guarded with secrecy for competitive advantage which in turn makes is susceptible to loss. Knowledge may also be lost from a catastrophic event -- such as the burning of the library of Alexandria~\cite{libalexandria}. Sometimes knowledge is lost out of neglect which may be benign or even malign -- benign neglect is one that simply loses the knowledge out of a  lack of an effort required in maintaining the knowledge; malign neglect on the other hand is a deliberate decision to stop such effort as is deemed to be incongruent with the prevailing ideals -- for example with a change in government, the new government may deliberately let old ideas die as it may not be aligned with the new ideals.

Sometimes we let go of a certain piece of knowledge because it is supplanted by something better that subsumes the earlier one. For instance with the invention of the transistors we could let go of vacuum tube technology (although we still know how to build them, we could in principle afford to forget it).  Another hypothetical example is replacing travel using ship by travel using airplanes for mass transit -- if we know how to efficiently transport everything by air we could in principle afford to let go of the knowledge of building great ships.  The knowledge that supplants another could be of a very different type -- the new knowledge lets us build new things that may be better,  but you won't be able to build the old things without the old  knowledge. Could this be considered as a `loss of knowledge'?

Sometimes we let go of knowledge because its charm seems  to fade -- it is not fashionable or cool anymore. For example in theoretical computer science computability theory is not as fashionable as it used to be whereas complexity theory is now. This is not necessarily for any apparent fundamental reason but simply a matter of culture. Similarly building tall skyscrapers was more fashionable in the 1970-80's than it is today.  Another example could be: in ancient times philosophy and religion was considered the most important type of knowledge; philosophers and religious leaders were considered the most respected scholars. Times have perhaps changed and today science has become the leading type of knowledge. So we have let go of  several of the old  religious and philosophical texts that were treasured as essential wisdoms in ancient times.  Perhaps it is because today we don't see that type of knowledge to be as useful as modern scientific knowledge. But fashions come and go in cycles~\cite{fashiontheory, fashionblog}. What is fashionable now may not be fashionable later and what was fashionable long ago may become fashionable in the future. Could it be the same with fashionable knowledge? Will we regret not having preserved knowledge that doesn't seem fashionable today.

But how can really useful knowledge be lost? Knowledge of technology that lets us move fast from one place to another, that lets us communicate, that improves our quality of life? Won't market forces ensure that these are always preserved or only replaced by better technology? Market forces preserve technology industries that are profitable, but they can also kill industries that may be useful but cease to be profitable (or cease to give the market rate of return). When market conditions change with evolution the cost of a certain technology may change significantly. In such cases a technology that earlier made economic sense for mass consumption may not anymore. This doesn't mean that the product that the technology creates is not useful anymore -- its just become more expensive to make. Then the market may force the technology to die and with it the knowledge of the technology begins to fade. Alternately, there may be several different unrelated but useful technologies competing for market resources. If there is no room for all the technologies due to shortage of resources (or public funds) then some technology may need to be put to rest. For a list of lost technologies see~\cite{top10} -- a classic example is the `Greek fire' a weapons technology based on fire that was lost with the decline of the Greek civilization. Knowledge can be lost and is lost, sometimes in large bursts, sometimes without us realizing it.

\vspace{0.2in}
\noindent{\bf The Need to simplify knowledge:}
What if we produce more and more knowledge that is useful? How will we ever maintain and preserve such a complex body of useful knowledge? Perhaps the only knowledge that the market forces will preserve is one that is concise. Perhaps if the knowledge of a technology becomes too complex, we will not be able to execute it economically unless we find a way to simplify the process of communicating and executing the knowledge. Necessity will force us to find efficient ways of expressing and storing that knowledge.  If we don't such knowledge may be unmanageable, unsearchable and may be lost.

\section{Forces that destroy and create knowledge}\label{forces}

What is the nature and strength of the forces that make us lose knowledge and 
that of forces that make us preserve it?

The force that makes us lose knowledge  is essentially the familiar force of entropy
underlying the second law of thermodynamics~\cite{thermoentropy}. This same force makes a
nice piece of glass eventually break up into pieces that it is almost next to
impossible to piece them together into their original form.  It is
the force that adds noise to bits, that moves things from order to disorder. This
force makes us forget things we know.
 This is also the force that given
a natural  source of random (or high entropy, affected by many diverse factors)
 bits, makes the bit 0 or 1 appear soon with very high probability. 

Although it is not a  physical forces like gravitation or electromagnetism, this
seemingly silent force is  one of the most powerful forces of
nature.
It is this force that brings decay and change. It is what makes us
lose things we had and sometimes accidentally
gives us new things out of the blue.
In many ways this force seems to make life difficult and frustrating; we cant
build a single simple solution for a problem that would work
everywhere -- every book needs to be translated to different languages, every machinery needs
to be adapted to the right voltage settings of the country, every cellphone needs to
understand the protocols used in the local industry -- everything seems to be in contention.
It is also perhaps this force of nature that creates such rich diversity -- so many
different languages, so many different wireless networking standards, so
many cultures and belief systems.

Our wish to preserve knowledge will need to fight this force or else the knowledge
will simply disappear under its mighty power. Some types of knowledge seems to
reappear again and again no matter how much it may seem to be lost. Like mathematics,
philosophy, religion, law -- as if they are ingrained in the human brain -- popping out on its
own -- rediscovering itself -- like a weed that cannot be killed. This too is a strong force --
one that preserves or rejuvenates knowledge.  This force creates knowledge. It is the force
that results in the strong human desire to seek knowledge, improve our lives, and to understand the world we live in. Some types
of knowledge are likely to be rediscovered sooner than others -- for example basic arithmetic is likely to
be rediscovered quickly even if society loses it, but something like group theory (or something
equivalent) may take a long time to be rediscovered if lost.

The force of entropy and the force of knowledge interact constantly resulting in rise and ebbs of
our knowledge. If we are ever able to conquer the force of entropy it must come from a knowledge that arises
from the force of knowledge.

\section{The Steady state of Knowledge. Could ever growing knowledge just be an Illusion?}\label{illusion}

If our knowledge as a society is only growing then what happens in the steady state? Is it the case that while our knowledge is growing on one hand, we
are unaware of another type of knowledge that is being lost -- some  knowledge we think is not interesting that we do not care to preserve. So in the steady
state perhaps we will constantly be producing more `apparently useful' knowledge not realizing a steady invisible process whereby another knowledge is getting lost -- may be even in
large bursts. In that case we would constantly be under the illusion that our knowledge is only growing. Could the present also be just one such state?

\section{Conclusions}
Preserving (important) knowledge is not easy. There are natural forces that destroy knowledge and there are forces that create knowledge. While we are aware of the new knowledge that is being constantly created we may be completely oblivious to the vast amounts of knowledge that is being forgotten -- we may forget things without realizing it. Thus it is easy to be in an illusion that knowledge is only increasing and not being lost.

\iftrue
\subsection*{Acknowledgements}
I would like to thank Sergey Yekhanin, Kostis Daskalakis, Madhu Sudan,  Marc Najork, Hari Sahasrabuddhe and Andrew Tomkins for useful discussions and comments.
\fi


\end{document}